# Three-dimensional graphene supported nickel molybdate nanowires as novel ultralight electrode for supercapacitors

Xiaozhi Liu[a,†], Ke Zhang[a,†], Baolin Yang[b], Wenlei Song[b], Qian Liu[b], Fei Jia[a], Shiyi Qin[a], Wanjun Chen[b], Zhenxing Zhang[b,*], Jian Li[b,*]

[a] Cuiying Honors College, Lanzhou University, Lanzhou 730000, P. R. China

[b] School of Physical Science and Technology, Lanzhou University, Lanzhou 730000, P. R. China

[†] These authors contributed equally to this work.

[*] Corresponding author. Tel.: +86 931 8912753; fax: +86 931 8913554. E-mail addresses: zhangzx@lzu.edu.cn (Z. Zhang), jianli@lzu.edu.cn (J. Li).

**ABSTRACT:** Nickel molybdate ($NiMoO_4$) nanowires were prepared on chemical-vapor-deposition-grown three-dimensional graphene skeletons by hydrothermal method. The X-ray diffraction and Raman results show $NiMoO_4$ nanowires are α phase. This binder-free and ultralight graphene/ $NiMoO_4$ composite was used as a positive electrode for supercapacitors. This electrode presents a high specific capacitance of 1194 F g$^{-1}$ at 12 mA cm$^{-2}$ and the good stability with a cycling efficiency of 97.3% after 1000 cycles. Further, the energy density reaches an energy density of 41 Wh kg$^{-1}$ at a steady power density of 1319 W kg$^{-1}$. These results demonstrate the potential of the designed composite for the future flexible and lightweight energy storage.

**Keywords:** Nickel molybdate; Graphene; Hydrothermal method; Carbon materials; Energy storage and conversion.

## 1. Introduction

Supercapacitors are an alternative environmentally friendly energy storage system with high-power and high-energy density, long lifespan, wide range of working temperature, and rapid recharging rate [1]. Nickel molybdate has been recently demonstrated to be a promising pseudocapacitor electrode material because of its high electrical conductivity, low cost, chemical and thermal stability, abundance and environment benignity. Nanoparticles [2], nanospheres [3], nanoporous [4], nanosheets [5, 6], nanowires [7, 8] of nickel molybdate for supercapacitors have been investigated. Although a high specific capacitance value of 2351 F $g^{-1}$ was reported [4], the used Ni foam current collector with large mass density (around 37 mg $cm^{-2}$) still seriously hinders electrochemical performance. Therefore, how to find an ultralight and binder-free current collector to improve $NiMoO_4$ electrodes' efficient energy storage remains a central challenge.

In this work, we used ultralight three-dimensional (3D) graphene skeletons as a substrate and current collector to grow $NiMoO_4$ nanowires by a simple one-pot hydrothermal method. The as-prepared 3D graphene/$NiMoO_4$ composite shows a high specific capacitance of 1194 F $g^{-1}$ at 12 mA $cm^{-2}$ and a good cycling stability of 97.3% after 1000 cycles.

2. **Material and methods**

First, 3D graphene skeletons were fabricated by chemical vapor deposition method as described previously [9]. The mass density of the obtained 3D graphene was 0.7 mg $cm^{-2}$. Then the 3D graphene/$NiMoO_4$ composite was prepared through a facile hydrothermal method. Briefly, 1.5 mmol of Ni $(NO_3)_2·6H_2O$, 1.5 mmol of $Na_2MoO_4·7H_2O$, and 30 mL deionized water were mixed and sealed in a 50-mL Teflon-lined autoclave. Then this autoclave was kept at 150 °C for 6 h and then cooled down to 30 °C naturally. The precursor was taken out and washed by distilled water and later acetone for a few seconds, then dried in a drying box at 80 °C for 12 h. Finally, the 3D graphene/$NiMoO_4$ composite was obtained by annealing the precursor at 300 °C for 1 h in argon gas. The mass loading of $NiMoO_4$ is 1.575 mg cm$^{-2}$.

The morphologies and crystal structures of the samples were characterized by field emission scanning electron microscopy (FE-SEM, Hitachi S-4800), X-ray diffraction (XRD, Philips X'pert pro, Cu Kα), and Raman spectroscopy (Jobin-Yvon HR 800), respectively. Electrochemical performance was tested on an electrochemical workstation (RST5200, Zhengzhou Shiruisi, China) with a three-electrode system. In tests, 2 M KOH aqueous solution was used as electrolyte, the 3D graphene/$NiMoO_4$ composite, a standard calomel electrode (SCE), and a Pt plate were used as working electrode, reference electrode, counter electrode, respectively.

3. **Results and discussion**

Fig. 1(a-b) shows the SEM images of graphene/$NiMoO_4$ composite. The nanowires are about 5 μm long and 120 nm in diameter, uniformly anchored on the 3D graphene skeletons. The labeled diffraction peaks in Fig. 1c correspond to α-$NiMoO_4$ phase (JCPDF#45-0142) and other peaks are from graphite phase (Fig. 1c). As shown in Fig. 1d, except for the characteristic G (1581 cm$^{-1}$) and 2D (2710 cm$^{-1}$) peaks from graphene skeleton, the Raman peaks from 100 to 1500 cm$^{-1}$ are from α-$NiMoO_4$ [9, 10]. These results demonstrate the graphene/$NiMoO_4$ composite is of high degree of purity and crystallinity.

Fig. 2 shows the asymmetrical cyclic voltammetry curves (CVs) of the 3D graphene/$NiMoO_4$ electrode, indicating the capacitive characteristics are mainly dominated by Faradaic reactions. Further, both the oxidation and the reduction currents are proportional to the square root of scan rate, implying the electrochemical reaction on the electrode surface is a diffusion-controlled process (as shown in the inset of Fig. 2a). Impressively, the specific capacitances of the whole composite, calculated from galvanostatic charge-discharge curves (GCDs), are 1194, 1023, 937, 869, 791, 673, and 549 F $g^{-1}$ at discharge current densities of 12, 24, 36, 54, 60, 90, and 120 mA $cm^{-2}$, respectively (as shown in Fig. 2c).

Fig. 2d shows the electrochemical impedance spectroscopy (EIS) measured in frequencies from 0.001 Hz to 100 kHz at open circuit voltage with an AC voltage perturbation amplitude of 50 mV. In the high-frequency region of the Nyquist plot, the characteristic impedance semicircle is mainly governed by the charge transferring process. Whereas, the tilting line in the low frequency zone is the feature of the electrolyte ions diffusion controlled kinetic process [11]. The EIS data has been fitted with an equivalent electrical circuit (as shown in the inset of Fig. 2d) with a minimized chi-squared ($\varkappa^2$) value in the order of $10^{-3}$, which consists of $R_S$ the resistance of the electrolyte combined with the internal resistance of the electrode, $R_{CT}$ the charge transfer resistance at the electrode –electrolyte interface, $W_O$ the Warburg type diffusion element, $CPE_{DL}$ the constant phase element representing double layer capacitance, and $CPE_L$ the pseudocapacitance, $R_L$ the leakage resistance. The intersecting point of the semicircle with the real axis gives the $R_S$ 1.59 Ohm which is consistent with the simulation result of 1.57 Ohm.

The power density (P, kW kg$^{-1}$) and energy density (E, Wh kg$^{-1}$), deduced from Equations $E = 0.5 \cdot C \cdot (\Delta V)^2$ and $P = E/t$ [9], are shown in Fig. 3a. At the current densities of 12 mA cm$^{-2}$, the resulting energy density is 41 Wh kg$^{-1}$ and the power density is 1194 W kg$^{-1}$, showing competitive performance to others. The electrochemical cycle stability test was performed at a current density of 48 mA cm$^{-2}$ in the range of 0-0.5 V (as shown in Fig. 3b). The initial specific capacitance is 869 F g$^{-1}$ and is decreased to 97% after 1000 cycles, which displays a long-term cycle stability of the 3D graphene/NiMoO$_4$ composite electrode. This should be attributed to the synergistic effect of graphene and NiMoO$_4$ nanowires.

4. **Conclusions**

In summary, we have successfully prepared NiMoO$_4$ nanowires on 3D graphene with outstanding pseudocapacitive performance for supercapacitors. The as-prepared electrode provides an excellent cyclic stability and a high value of specific capacitance about 1194 F g$^{-1}$ at 12 mA cm$^{-2}$ with respect to the total mass of the electrode. Our work confirms the 3D graphene is a superior current collector for the future ultralight supercapacitors, and illuminates a new way to utilize the capacitive performance of the electroactive materials.

**Acknowledgments**

This research was supported by the National Natural Science Foundation of China (No.: 51302122), the Natural Science Foundation of Gansu Province in China (No:

**Figure 1**

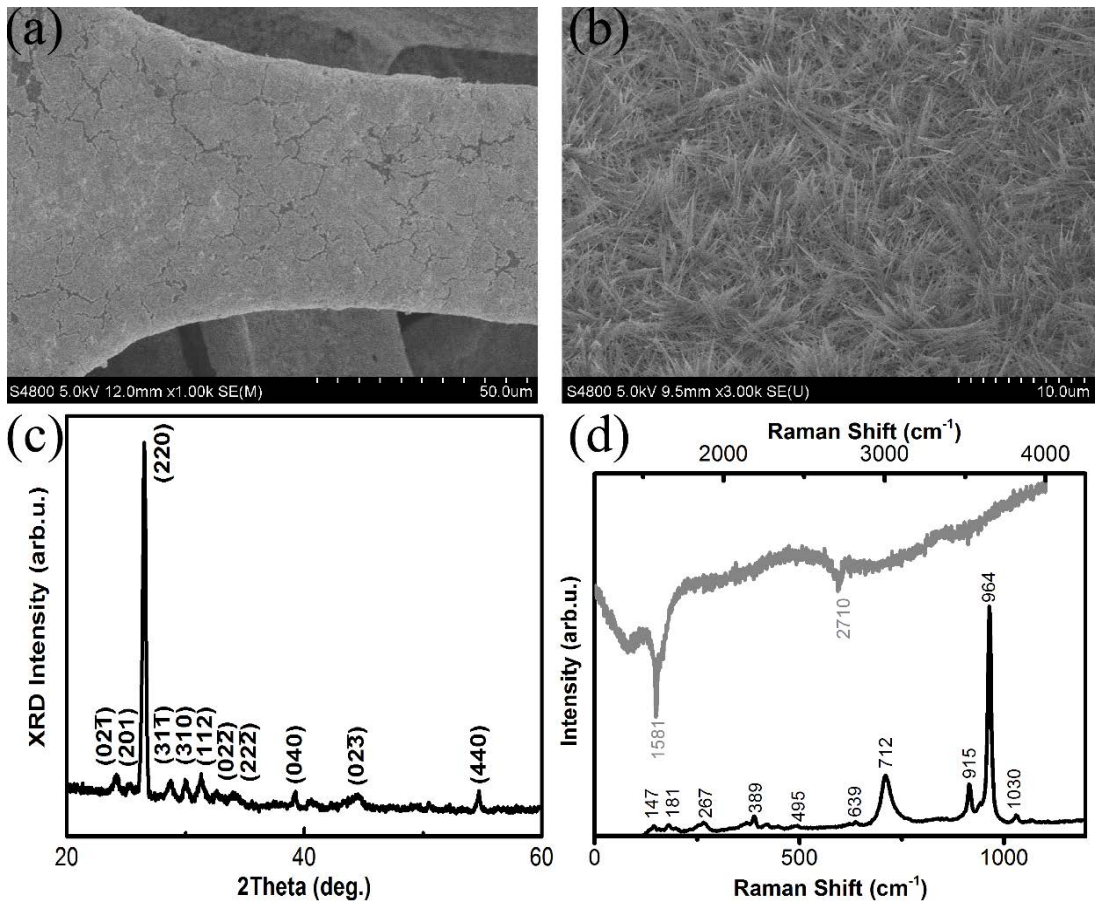

**Figure 2**

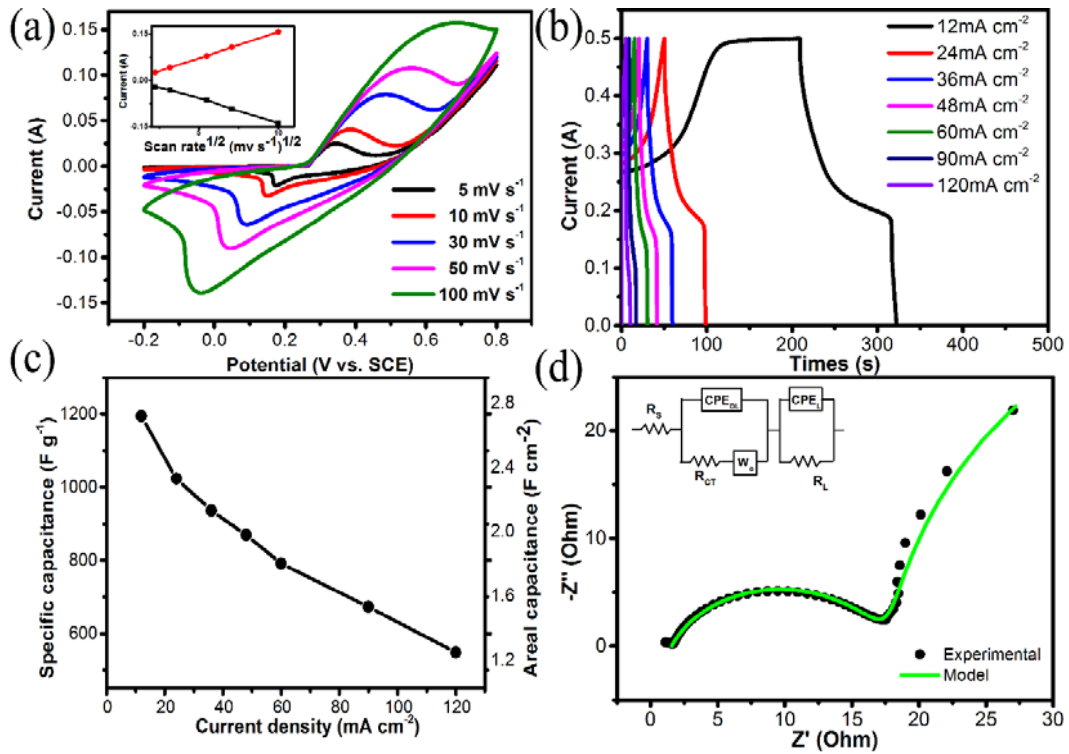

**Figure 3**

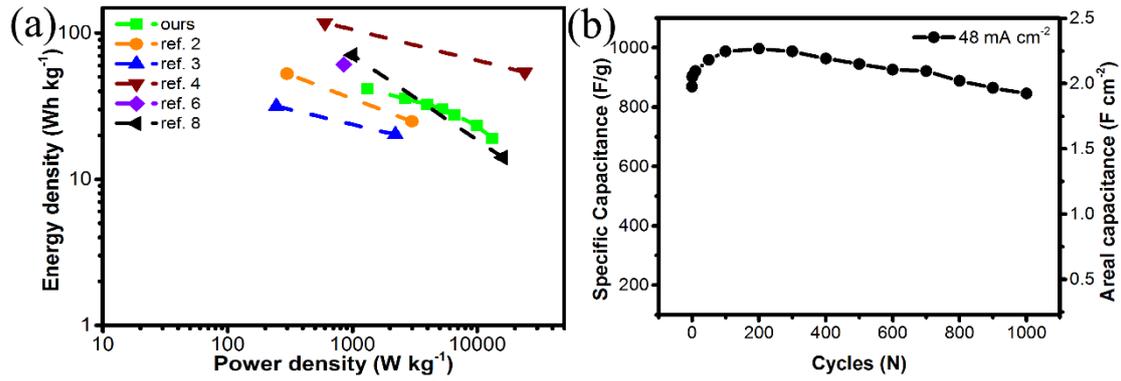

**Figure captions:**

Figure 1 (a) SEM images of graphene/NiMoO$_4$ nanowire skeletons and (b) NiMoO$_4$ nanowires; (c) XRD pattern of graphene/NiMoO$_4$, and (d) Raman spectra of graphene and graphene/NiMoO$_4$.

Figure 2 (a) CVs of graphene/NiMoO$_4$ at different scan rate. The inset is the plots of oxidation peak current and reduction peak current vs. the square root of the scan rate; (b) GCDs of graphene/NiMoO$_4$; (c) Areal and specific capacitance vs. current density for graphene/NiMoO$_4$; (d) EIS plot of graphene/NiMoO$_4$.

Figure 3 (a) Ragone plots of the graphene/NiMoO$_4$; (b) Cycling performance of graphene/NiMoO$_4$ at a current density of 48 mA cm$^{-2}$.